\documentclass[aps,prd,onecolumn,groupedaddress,showpacs,nofootinbib,amssymb]{revtex4}
\usepackage[dvips]{graphicx}
\usepackage{amssymb}
\usepackage{amsmath}
\usepackage{graphicx,,color}
\usepackage{amsfonts}
\usepackage{bm}
\usepackage{cancel}
\usepackage{comment}

\newcommand\be{\begin{equation}}
\newcommand\ee{\end{equation}}

\allowdisplaybreaks[4]

\begin{document}

\title{A Study of an Einstein Gauss-Bonnet Quintessential Inflationary Model}
\author{
K. Kleidis,$^{1}$\,\thanks{kleidis@teiser.gr} V.K.
Oikonomou$^{2,1}$,\,\thanks{v.k.oikonomou1979@gmail.com}}
\affiliation{ $^{1)}$ Department of Mechanical Engineering\\
Technological
Education Institute of Central Macedonia \\
62124 Serres, Greece \\
$^{2)}$ Department of Physics, Aristotle University of
Thessaloniki, Thessaloniki 54124, Greece }

\tolerance=5000

\begin{abstract}
In this paper we study a class of quintessential Einstein
Gauss-Bonnet models, focusing on their early and late-time
phenomenology. With regard to the early-time phenomenology, we
formalize the slow-roll evolution of these models and we calculate
in detail the spectral index of the primordial curvature
perturbations and the tensor-to-scalar ratio. As we demonstrate,
the resulting observational indices can be compatible with both
the Planck and the BICEP2/Keck-Array observational constraints on
inflation. With regard to the late-time behavior, by performing a
numerical analysis we demonstrate that the class of models for
which the coupling function $\xi (\phi)$ to the Gauss-Bonnet
scalar satisfies $\xi (\phi)\sim \frac{1}{V(\phi )}$, produce a
similar pattern of evolution, which at late-times is characterized
by a decelerating era until some critical redshift, at which point
the Universe super-decelerates and subsequently accelerates until
present time, with a decreasing rate though. The critical redshift
crucially depends on the initial conditions chosen for the scalar
field and for all the quintessential Einstein Gauss-Bonnet models
studied, the late-time era is realized for large values of the
scalar field.
\end{abstract}


\maketitle

\section{Introduction}

The early-time acceleration era of our Universe is the last resort
of the classical physics to our Universe's description. Prior to
the early acceleration era, it is believed that strong gravity
effects take control of the physical description, thus making it
inaccessible to our moderate classical description. The early-time
acceleration era dubbed inflationary era, can be described by a
classical theory of gravity and also can be constrained by the
observations, since the primordial modes after these exit the
horizon, can be relevant to present day observations if these
freeze or evolve slowly before they reenter the horizon during the
radiation domination era. From the introduction of the early
inflationary models in the early 80's
\cite{Guth:1980zm,Linde:1993cn,Linde:1983gd}, major progress has
been achieved, especially after the Planck observational data
\cite{Ade:2015lrj} have been released. Particularly, the
observational constraints have reduced significantly the number of
viable inflationary models. Apart from the standard inflationary
paradigm of a slow-rolling canonical scalar field, there exist
also alternative scenarios that can also provide a viable
inflationary era
\cite{inflation1,inflation2,inflation3,inflation4,reviews1,reviews2},
see also \cite{reviews3,reviews4,reviews5,reviews6}. The vital
question is, which description can be the correct description of
our Universe. This question is not easy to answer, and for the
time being, no single-model answer can be given. Therefore, it is
vital to investigate several scenarios that can yield a viable
inflationary era, thus covering all possible answers to the
question which model is the best description that fits the
observational data. Apart from the early-time era, a viable model
must also describe the late-time era of our Universe, which
currently is expanding in an accelerating way. One appealing class
of models is the so-called quintessential inflation models
\cite{Dimopoulos:2017tud,vandeBruck:2017voa,AresteSalo:2017lkv,Agarwal:2017wxo,Ahmad:2017itq,Geng:2017mic,Dimopoulos:2017zvq,deHaro:2016ftq,Haro:2015ljc,Hossain:2014zma,Hossain:2014coa,Neupane:2007mu,Sami:2004xk,Tashiro:2003qp,Dimopoulos:2002hm,Dimopoulos:2001ix,Peebles:1998qn},
according to which both the early and late-time acceleration eras
are consistently described. In this paper we shall consider an
Einstein Gauss-Bonnet extension of a specific quintessential model
studied in Ref. \cite{Geng:2017mic}. The motivation for the study
of the Einstein Gauss-Bonnet extension comes from the fact that
the primordial accelerating era can potentially be affected by the
preceding strong gravity era, thus it is possible that string
theory effects can have an imprint on the classically evolving
canonical scalar theory which can describe the inflationary
Universe. In the literature, there exist various studies of this
sort
\cite{Cognola:2006sp,Nojiri:2005vv,Nojiri:2005jg,Nojiri:2007te,Satoh:2008ck,Satoh:2007gn,Hikmawan:2015rze,Bamba:2014zoa,Yi:2018gse,Guo:2009uk,Guo:2010jr,Jiang:2013gza,Koh:2014bka,Koh:2016abf,Kanti:2015pda,vandeBruck:2017voa,Kanti:1998jd,Nozari:2017rta,Chakraborty:2018scm,Odintsov:2018zhw},
and in this paper we will critically investigate the
quintessential inflation scenario. Our main aim is to investigate
whether a viable primordial accelerating era can be produced, and
secondly, whether a late-time accelerating era can also be
produced. With regard to the latter, we will investigate the
phenomenological features of the theory, mainly focusing on the
behavior of the deceleration parameter and of the total effective
equation of state parameter. The results of our study indicate
that a viable inflationary era, compatible with the Planck
\cite{Ade:2015lrj} and the BICEP2/Keck-Array data
\cite{Array:2015xqh}, can easily be produced for all the Einstein
Gauss-Bonnet quintessential models which we shall study. In fact,
the rich parameter space enhances the viability of the single
scalar field quintessential inflation scenario. In addition, as we
will show, it is possible to obtain a late-time accelerating era,
with the transition from deceleration to acceleration depending
however strongly on the initial conditions chosen for the scalar
field. As we will demonstrate, the late-time acceleration era is
different for the models studied, in comparison to the
$\Lambda$-Cold-Dark-Matter ($\Lambda$CDM) model, and there seems
to be a pattern of common behavior for models for which the
coupling function to the Gauss-Bonnet scalar $\xi (\phi)$
satisfies the relation $\xi (\phi)\sim \frac{1}{V(\phi)}$.

This paper is organized as follows: In section II we formulate the
slow-roll dynamics of the Einstein Gauss-Bonnet quintessential
inflation theory. Moreover, we investigate the viability of two
models, by calculating in detail the spectral index of the
primordial curvature perturbations and the tensor-to-scalar ratio
and finally by confronting the results with the latest Planck
constraints. In section III we numerically study the late-time
behavior of the model by expressing the gravitational equations of
motion as functions of the redshift parameter $z$, and focusing on
values of $z$ in the range $z=[0,10]$. We quantify our study by
examining the behavior of the deceleration parameter $q(z)$ and of
the total effective equation of state parameter $w_{eff}(z)$ as
functions of the redshift. Finally, the conclusions follow at the
end of the paper.

Prior proceeding, we need to note that the geometric background
which will be assumed in this paper is a flat
Friedmann-Robertson-Walker (FRW) metric, with line element,
\begin{equation}
\label{metricfrw} ds^2 = - dt^2 + a(t)^2 \sum_{i=1,2,3}
\left(dx^i\right)^2\, ,
\end{equation}
where $a(t)$ is the scale factor. Also, the metric connection is
assumed to be the Levi-Civita connection.

\section{Slow-roll Dynamics of Einstein Gauss-Bonnet Gravity and Quintessential Inflation}

The quintessential inflation scenario is appealing by itself since
it is possible to describe in a unified way a viable inflationary
era compatible with the observational data, and an accelerating
late-time evolution with the total effective equation of state
parameter satisfying $w_{eff}<-\frac{1}{3}$. In this section we
shall present the general Einstein Gauss-Bonnet modification of
the quintessential inflation scenario, and we investigate the
effects of the Gauss-Bonnet coupling $\xi(\phi)\mathcal{G}$ to the
canonical scalar field quintessential scenario. As we will
demonstrate, the parameter space for which the viability with
observations is achieved is enlarged in the Einstein Gauss-Bonnet
case.

So we assume that the general $f(\phi,R)$ Einstein Gauss-Bonnet
gravity that controls the Universe's evolution, has the following
action \cite{Bamba:2014zoa},
\begin{equation}\label{actionegbgeneral}
\mathcal{S}=\int
\mathrm{d}^4x\sqrt{-g}\left(\frac{1}{2}f(\phi,R)-\frac{\omega(\phi
)}{2}\partial_{\mu}\phi\partial^{\mu}\phi-V(\phi)-\frac{c_1}{2}\xi
(\phi )\mathcal{G} \right)\, ,
\end{equation}
where $\mathcal{G}$ stands for the Gauss-Bonnet scalar, which in
terms of the Ricci scalar, the Ricci tensor and the Riemann tensor
is written as follows,
\begin{equation}\label{gaussbonnetscalar}
\mathcal{G}=R^2-4R_{\mu \nu}R^{\mu \nu}+R_{\mu \nu \sigma
\rho}R^{\mu \nu \sigma \rho}\, ,
\end{equation}
which for the FRW metric (\ref{metricfrw}) it takes the form
$\mathcal{G}=24 (H^2+\dot{H})$. In Eq. (\ref{actionegbgeneral}) we
chose the reduced Planck units system, for which $\hbar=c=1$ and
also $\kappa^2=\frac{8\pi}{M_p}^2=1$, where $M_p$ is the Planck
mass scale. In the following sections we shall consider the case
$f(R,\phi)=R$ and $\omega (\phi)=1$, but for the sake of
generality we shall present the equations of motion with general
forms of the aforementioned functions. Upon variation of the
action (\ref{actionegbgeneral}) with respect to the metric tensor
$g_{\mu \nu}$, the equations of motion are obtained, which read,
\begin{align}\label{eqnsofmotionfrw}
& \frac{\omega
(\phi)}{2}\dot{\phi}^2+V(\phi)+\frac{R}{2}F-\frac{f(\phi,R)}{2}-3F(\phi,R)H^2+12c_1\xi'(\phi)\dot{\phi}H^3=0\,
,\\ \notag & \frac{\omega
(\phi)}{2}\dot{\phi}^2-V(\phi)+\frac{f(\phi,R)}{2}-3F(\phi,R)(\dot{H}+3H^2)+2\dot{F}H+\ddot{F}-4c_1\left(
H^2\dot{\phi}^2\xi''(\phi)+H^2\ddot{\phi}\xi'(\phi)+2H(\dot{H}+H^2)\dot{\phi}\xi'(\phi)\right)=0\,
,
\end{align}
and moreover the variation of the action with respect to the
scalar field yields the following equation,
\begin{equation}\label{varwiscalar}
\omega (\phi)\ddot{\phi}+3\omega
(\phi)H\dot{\phi}V'(\phi)+\frac{1}{2}\omega
'(\phi)-\frac{f'(\phi,R)}{2}+12c_1\xi'(\phi)H^2(\dot{H}+H^2)=0\, ,
\end{equation}
where the ``prime'' indicates differentiation with respect to
$\phi$, while the ``dot'' with respect to the cosmic time. Also
the function $F$ in the above equations is equal to
$F=\frac{\partial f(\phi,R)}{\partial R}$.

\begin{figure}[h]
\centering
\includegraphics[width=18pc]{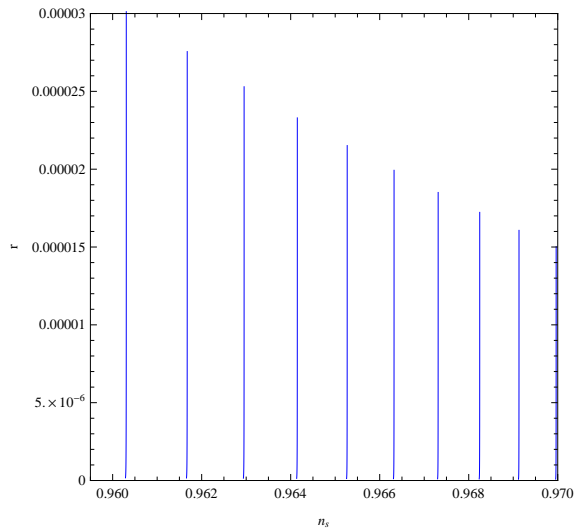}
\caption{Parametric plot of the spectral index and of the
tensor-to-scalar ratio as a function of the parameters $c_1$ and
$\beta$, for the Einstein Gauss-Bonnet theory with $\xi (\phi)=V_0
e^{\beta  \phi ^3}$ and $V (\phi)=V_0 e^{-\beta  \phi ^3}$, with
$N=60$ and $V_0=10$. The different lines correspond to various
values of the parameters $c_1$ and $\beta$ in the ranges
$c_1=[0.03,0.05]$ with step 0.0009 and $\beta=[0.01,0.13]$. The
plots correspond to allowed values of $c_1$ and $\beta$, for which
the spectral index and the tensor-to-scalar ratio are
simultaneously compatible with the observational
data.}\label{plot1}
\end{figure}

The inflationary dynamics for a generalized $f(R,\phi)$ Einstein
Gauss-Bonnet theory were studied in
\cite{Noh:2001ia,Hwang:2005hb,Hwang:2002fp}, according to which
the slow-roll indices for the action (\ref{actionegbgeneral}) are
equal to,
\begin{equation}\label{slowrollindicesini}
\epsilon_1=\frac{\dot{H}}{H^2},\,\,\,\epsilon_2=\frac{\ddot{\phi}}{H\dot{\phi}},\,\,\,\epsilon_3=\frac{\dot{F}}{2HF},\,\,\,\epsilon_4=\frac{\dot{E}}{2HE},\,\,
\,\epsilon_5=\frac{\dot{F}+Q_a}{H(2F+Q_b)},\,\,\,\epsilon_6=\frac{\dot{Q}_t}{2HQ_t}\,
,
\end{equation}
with the function $E$ being defined in the following way,
\begin{equation}\label{epsiloncapdefinit}
E=\frac{F}{\dot{\phi}}\left( \omega
(\phi)\dot{\phi}^2+3\frac{(\dot{F}+Q_a)^2}{2F+Q_b}\right)\, ,
\end{equation}
and in addition, the functions $Q_a$, $Q_b$ and $Q_t$ are equal
to,
\begin{equation}\label{qaqb}
Q_a=-4c_1\dot{\xi}H^2,\,\,\,Q_b=-8c_1\dot{\xi} H, \,\,\,
Q_t=F+\frac{1}{2}Q_b\, .
\end{equation}
We shall assume that the slow-roll conditions hold true for the
theory at hand, so by assuming the condition $\epsilon_i\ll 1$,
$i=1,2,..,6$ for the slow-roll indices, we obtain the following
slow-roll conditions that must be satisfied by the Hubble rate and
by the functions $\xi (\phi)$ and $V(\phi)$,
\begin{equation}\label{slowrollcond1}
\dot{H}\ll H^2,\,\,\,\ddot{\phi}\ll
H\dot{\phi},\,\,\,\dot{\xi}H\ll 1,\,\,\,\ddot{\xi}\ll
1,\,\,\,\dot{\xi}\dot{H}\ll 1\, .
\end{equation}
where we used the fact that we are considering a theory with
$f(R,\phi )=R$, $F(\phi,R)=1$ and $f'(\phi,R)=0$. Since the
slow-roll conditions for the scalar field $\phi$ imply that
$3H^2\sim V(\phi)$, from the equations of motion we obtain the
following two equations for $\dot{\phi}$ and $\dot{H}$,
\begin{align}\label{derivativesofphiandh}
&
\dot{\phi}=-\frac{12c_1\xi'(\phi)V(\phi)^2}{27\sqrt{\frac{V(\phi)}{3}}}-\frac{V'(\phi)}{3\sqrt{\frac{V(\phi)}{3}}},
\\ \notag & \dot{H}=4c_1H^3\xi'(\phi)-\frac{\dot{\phi}^2}{4}\, .
\end{align}
In view of the above equations, the slow-roll indices can be
written as follows,
\begin{align}\label{slowrollindicesasfunctionsofphi}
& \epsilon_1=-\frac{V'(\phi)^2}{4V(\phi)^2},\\ \notag &
\epsilon_2=2\frac{V''(\phi)}{V(\phi)},\\ \notag & \epsilon_3=0,\\
\notag & \epsilon_4=-\frac{4 c_1^3 \sqrt{V(\phi )} \xi '(\phi )^3
\left(4 c_1 V(\phi )^2 \xi '(\phi )+3 V'(\phi )\right) \left(20
c_1 V(\phi )^2 \xi '(\phi )+3 V'(\phi )\right)}{27 \sqrt{3}}\, .
\end{align}
As it was shown in \cite{Noh:2001ia,Hwang:2005hb,Hwang:2002fp},
the spectral index of the primordial curvature perturbations and
the tensor-to-scalar ratio for the theory at hand in the slow-roll
approximation, are given below,
\begin{equation}\label{observationalindices1}
n_s\simeq=1+4\epsilon_1-2\epsilon_2+2\epsilon_3-2\epsilon_4,\,\,\,r=4\left(\epsilon_1-\frac{1}{4}(-\frac{Q_e(t)}{H}+Q_f(t))
\right)\, ,
\end{equation}
with $Q_e$ and $Q_f$ being equal to,
\begin{equation}\label{qeqf}
Q_e=8c_1\dot{\xi}\dot{H},\,\,\,Q_f=-4c_1(\ddot{\xi}-\dot{\xi}H)\,
.
\end{equation}
From Eq. (\ref{derivativesofphiandh}), in conjunction with Eqs.
(\ref{slowrollindicesasfunctionsofphi}),
(\ref{observationalindices1}) and (\ref{qeqf}), the observational
indices read,
\begin{align}\label{observationalindicesfinalforms}
& n_s\simeq 1-\frac{8 c_1^3 \sqrt{V(\phi )} \xi '(\phi )^3 \left(4
c_1 V(\phi )^2 \xi '(\phi )-3 V'(\phi )\right) \left(4 c_1 V(\phi
)^2 \xi '(\phi )+V'(\phi )\right)}{9 \sqrt{3}}-\frac{4 V''(\phi
)}{V(\phi )}-\frac{V'(\phi )^2}{V(\phi )^2},\\ \notag & r\simeq
\Big{|}-\frac{32}{9} c_1^2 V(\phi )^2 \xi '(\phi )^2-\frac{8}{3}
c_1 \xi '(\phi ) V'(\phi )-\frac{2 V'(\phi )^2}{V(\phi
)^2}\Big{|}\, .
\end{align}
Let us now exemplify how the above formalism can be used in the
case of some models of quintessential inflation. The purpose of
this section is to confront certain classes of Einstein
Gauss-Bonnet quintessential inflation models with the latest 2015
Planck data \cite{Ade:2015lrj} and with the BICEP2/Keck-Array data
\cite{Array:2015xqh}, which constrain the spectral index $n_s$ and
the tensor-to-scalar ratio $r$ as follows, \cite{Array:2015xqh},
\begin{equation}
\label{planckdata} n_s=0.9644\pm 0.0049\, , \quad
r<0.10\,\,\,(\mathrm{Planck}\,\,\,2015)\, ,
\end{equation}
\begin{equation}
\label{scalartotensorbicep2}
r<0.07\,\,\,(\mathrm{BICEP2}/\mathrm{Keck-Array})\, .
\end{equation}
We shall firstly consider the following scalar potential,
\begin{equation}\label{quintpot1}
V(\phi)=V_0e^{-\beta \phi^3}\, ,
\end{equation}
where $\beta$ and $V_0$ are arbitrary positive and real numbers.
Also we assume that the Gauss-Bonnet coupling function $\xi
(\phi)$ has the following form,
\begin{equation}\label{xiphicoupling1}
\xi (\phi)=V_0e^{\beta \phi^3}\, ,
\end{equation}
hence the coupling function $\xi (\phi)$ and the potential
$V(\phi)$ satisfy $\xi(\phi)\sim \frac{1}{V(\phi)}$. In a later
section we shall demonstrate that this general class of models
produces an accelerating late-time era, quite similar for all the
models that belong to this class. The slow-roll indices as a
function of the scalar field read,
\begin{equation}\label{slowrollindicescase1new1}
\epsilon_1\simeq -\frac{9}{4} \beta ^2 \phi
^4,\,\,\,\epsilon_2\simeq 6 \beta  \phi  \left(3 \beta  \phi
^3-2\right),\,\,\,\epsilon_4\simeq 36 \sqrt{3} \beta ^5 c_1^3
V_0^{11/2} \phi ^{10} e^{\frac{\beta  \phi ^3}{2}} \left(4 c_1
V_0^2-1\right) \left(4 c_1 V_0^2+3\right)\, ,
\end{equation}
and the observational indices for inflation are equal to,
\begin{align}\label{obsevrationalindciesexample1new1}
& n_s\simeq 1-45 \beta ^2 \phi ^4+24 \beta  \phi -72 \sqrt{3}
\beta ^5 c_1^3 V_0^{11/2} \phi ^{10} e^{\frac{\beta  \phi ^3}{2}}
\left(16 c_1^2 V_0^4+8 c_1 V_0^2-3\right),
\\ \notag & r\simeq \Big{|}-18 \beta ^2 \phi ^4+32 \beta ^2 c_1^2 V_0^4 \phi ^4+24 \beta ^2 c_1 V_0^2 \phi ^4\Big{|}\, .
\end{align}
The functional form of the slow-roll indices indicates that the
slow-roll inflationary era can be realized for small values of the
scalar field, that is for $\phi\ll 1$. It is worthy expressing the
observational indices as function of the $e$-foldings number,
which is defined as  a function of the Hubble rate as follows,
\begin{equation}\label{efoldingsdefinition1}
N=\int_{t_i}^{t_f}H(t)\mathrm{d}t\, ,
\end{equation}
with $t_i$, $t_f$ being the time instance of the beginning and end
of inflation respectively. By expressing the $e$-foldings as a
function of the scalar field for the slow-roll Einstein
Gauss-Bonnet theory, we obtain the following formula,
\begin{equation}\label{efoldignsscalarfield}
N\simeq \int_{\phi_k}^{\phi_f}-\frac{3 V(\phi )}{4 c_1 V(\phi )^2
\xi '(\phi )+3 V'(\phi )}\, ,
\end{equation}
with $\phi_k$ and $\phi_f$ being the scalar field values at the
horizon crossing and at the end of inflation respectively. The
value $\phi_f$ can be determined by the condition
$|\epsilon_1(\phi_f)|\simeq \mathcal{O}(1)$, so we have
$\phi_f\simeq \sqrt{\frac{2}{3 \beta }}$. The value of the scalar
field at the horizon crossing can be found by using Eq.
(\ref{efoldignsscalarfield}), so the resulting $\phi_k$ is,
\begin{equation}\label{phikexample1}
\phi_k\simeq \frac{2}{\beta  \left(\sqrt{6} \sqrt{\beta }-8 c_1 N
V_0^2+6 N\right)}\, .
\end{equation}
Now calculating the observational indices of inflation at the
horizon crossing instance, that is at $\phi=\phi_k$, and by using
Eq. (\ref{phikexample1}), we can express $n_s$ and $r$ as
functions of the $e$-foldings number, so we get,
\begin{align}\label{nsnexample1new1}
& n_s\simeq 1-\frac{720}{\beta ^2 \left(\sqrt{6} \sqrt{\beta }-8
c_1 N V_0^2+6 N\right)^4}+\frac{48}{\sqrt{6} \sqrt{\beta }-8 c_1 N
V_0^2+6 N}\\ \notag & -\frac{73728 \sqrt{3} c_1^3 V_0^{11/2}
\left(16 c_1^2 V_0^4+8 c_1 V_0^2-3\right) e^{\frac{4}{\beta ^2
\left(\sqrt{6} \sqrt{\beta }+N \left(6-8 c_1
V_0^2\right)\right)^3}}}{\beta ^5 \left(\sqrt{6} \sqrt{\beta }+N
\left(6-8 c_1 V_0^2\right)\right)^{10}}\, ,
\end{align}
\begin{equation}\label{scalartotensorrationexample1new1}
r\simeq \Big{|} \frac{32 \left(16 c_1^2 V_0^4+12 c_1
V_0^2-9\right)}{\beta ^2 \left(\sqrt{6} \sqrt{\beta }+N \left(6-8
c_1 V_0^2\right)\right)^4} \Big{|}\, .
\end{equation}
Having the functional form of the observational indices as
functions of the $e$-foldings and of the parameters given in Eqs.
(\ref{nsnexample1new1}) and
(\ref{scalartotensorrationexample1new1}), we can directly confront
the Einstein Gauss-Bonnet theory
(\ref{quintpot1})-(\ref{xiphicoupling1}) with the observational
data (\ref{planckdata}) and (\ref{scalartotensorbicep2}). A
thorough analysis indicates that the parameter space is quite
large and it allows the theory to be compatible with the
observational data for a wide range of the parameter values. For
example by choosing $V_0\sim \mathcal{O}(10)$, $N=60$ and
$\beta=0.01$, $c_1=0.04$, we obtain $n_s\simeq 0.969225$ and
$r\simeq 0.0000159495$, which are both compatible with the
observational data. Also the canonical scalar field theory can
also be compatible with the observational data, for example if
$N=60$, $V_0\sim \mathcal{O}(10)$ and for $c_1=0$ and
$\beta=0.0685454$, we get $n_s=0.96195$ and $r=0.06$ which are
also compatible with the observational constraints
(\ref{planckdata}) and (\ref{scalartotensorbicep2}). Therefore,
the Einstein Gauss-Bonnet theory of quintessential inflation
enlarges the range of parameter values which render the model
compatible with the observational data. This can also be seen in
Fig. \ref{plot1}, where we have presented the parametric plot of
the spectral index and of the tensor-to-scalar ratio as a function
of the parameter $\beta$, for $N=60$, $V_0=10$ and for
$c_1=[0.03,0.05]$ with step 0.0009 and $\beta=[0.01,0.13]$.

\begin{figure}[h]
\centering
\includegraphics[width=18pc]{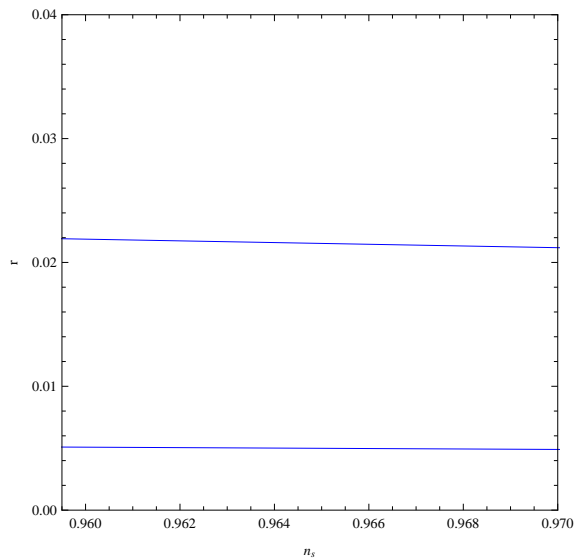}
\caption{Parametric plot of the spectral index and of the
tensor-to-scalar ratio as a function of the parameters $c_1$ and
$\beta$, for the Einstein Gauss-Bonnet theory with $\xi (\phi)=V_0
e^{\beta  \phi ^4}$ and $V (\phi)=V_0 e^{-\beta  \phi ^4}$, with
$N=60$ and $V_0=7$. The different lines correspond to various
values of the parameters $c_1$ and $\beta$ in the ranges
$c_1=[0.008,0.05]$ with step 0.0001 and $\beta=[8\times
10^{-6},0.01]$. The plots correspond to allowed values of $c_1$
and $\beta$, for which the spectral index and the tensor-to-scalar
ratio are simultaneously compatible with the observational
data.}\label{plot2}
\end{figure}

Consider now the case that $V(\phi)$ and $\xi (\phi)$ are assumed
to be,
\begin{equation}\label{potentialxithrirdnew2}
V(\phi)=V_0e^{-\beta \phi^4},\,\,\,\xi (\phi)=V_0e^{\beta
\phi^4}\, ,
\end{equation}
In this case, the slow-roll indices read,
\begin{equation}\label{slowrollindicescase1new2}
\epsilon_1\simeq -4 \beta ^2 \phi ^6,\,\,\,\epsilon_2\simeq 8
\beta  \phi ^2 \left(4 \beta  \phi
^4-3\right),\,\,\,\epsilon_4\simeq \frac{4096 \beta ^5 c_1^3
V_0^{11/2} \phi ^{15} e^{\frac{\beta  \phi ^4}{2}} \left(4 c_1
V_0^2-1\right) \left(4 c_1 V_0^2+3\right)}{9 \sqrt{3}}\, ,
\end{equation}
and the observational indices are equal to,
\begin{align}\label{obsevrationalindciesexample1new2}
& n_s\simeq \frac{1}{27} \left(-2160 \beta ^2 \phi ^6+1296 \beta
\phi ^2-8192 \sqrt{3} \beta ^5 c_1^3 V_0^{11/2} \phi ^{15}
e^{\frac{\beta  \phi ^4}{2}} \left(16 c_1^2 V_0^4+8 c_1
V_0^2-3\right)+27\right),
\\ \notag & r\simeq \Big{|}-32 \beta ^2 \phi ^6+\frac{512}{9} \beta ^2 c_1^2 V_0^4 \phi ^6+\frac{128}{3} \beta ^2 c_1 V_0^2 \phi ^6\Big{|}\, .
\end{align}
From the functional form of the slow-roll indices as functions of
the scalar field $\phi$, namely Eq.
(\ref{slowrollindicescase1new2}), it is obvious that the slow-roll
era is realized for small values of the scalar field. Following
the procedure of the previous case, the observational indices as
functions of the $e$-foldings number are,
\begin{align}\label{nsnexample1new2}
&n_s\simeq 1\frac{144 \beta }{3\ 2^{2/3} \beta ^{2/3}-8 \beta  N
\left(4 c_1 V_0^2-3\right)}-\frac{2160}{\left(3\ 2^{2/3}-8
\sqrt[3]{\beta } N \left(4 c_1 V_0^2-3\right)\right)^3}\\ \notag &
+\frac{\left(-31850496 c_1^5 V_0^{10}-15925248 c_1^4 V_0^8+5971968
c_1^3 V_0^6\right) \exp \left(\frac{9 \beta }{2 \left(3\ 2^{2/3}
\beta ^{2/3}-8 \beta  N \left(4 c_1
V_0^2-3\right)\right)^2}\right)}{\sqrt{V_0} \left(3\ 2^{2/3}-8
\sqrt[3]{\beta } N \left(4 c_1 V_0^2-3\right)\right)^{15/2}}\, ,
\end{align}
\begin{equation}\label{scalartotensorrationexample1new2}
r\simeq \Big{|} \frac{96 \left(16 c_1^2 V_0^4+12 c_1
V_0^2-9\right)}{\left(3\ 2^{2/3}-8 \sqrt[3]{\beta } N \left(4 c_1
V_0^2-3\right)\right)^3} \Big{|}\, .
\end{equation}
As in the previous case, the model (\ref{potentialxithrirdnew2})
can also be compatible with the observational constraints on
inflation, for a wide range of parameter values. In fact, the
viability of the canonical scalar quintessential inflation model
is enlarged. In Fig. \ref{plot2} we present the parametric plot of
the spectral index and of the tensor-to-scalar ratio as a function
of the parameters $c_1$ and $\beta$, with $N=60$ and $V_0=7$. The
different lines correspond to various values of the parameters
$c_1$ and $\beta$ in the ranges $c_1=[0.008,0.05]$ with step
0.0001 and $\beta=[8\times 10^{-6},0.01]$.
Now having discussed the inflationary properties of the Einstein
Gauss-Bonnet extended quintessential inflationary models, what
remains is to examine the late-time properties. The canonical
scalar field quintessential model describes an accelerating
late-time evolution, so it is vital to investigate the late-time
behavior of the Einstein Gauss-Bonnet extensions we discussed
earlier. This is the subject of the next section.

\section{Late-time Evolution of Quintessential Einstein Gauss-Bonnet Models}

\begin{figure}[h]
\centering
\includegraphics[width=18pc]{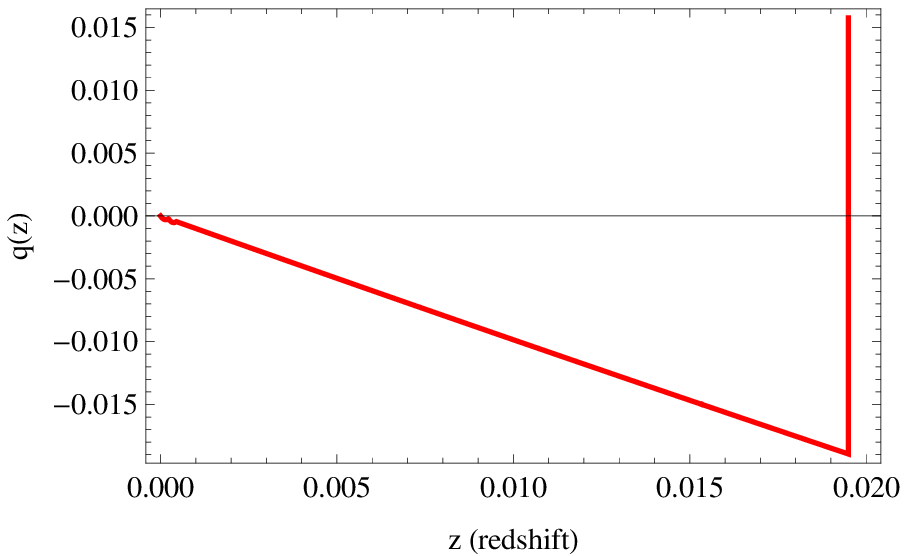}
\includegraphics[width=18pc]{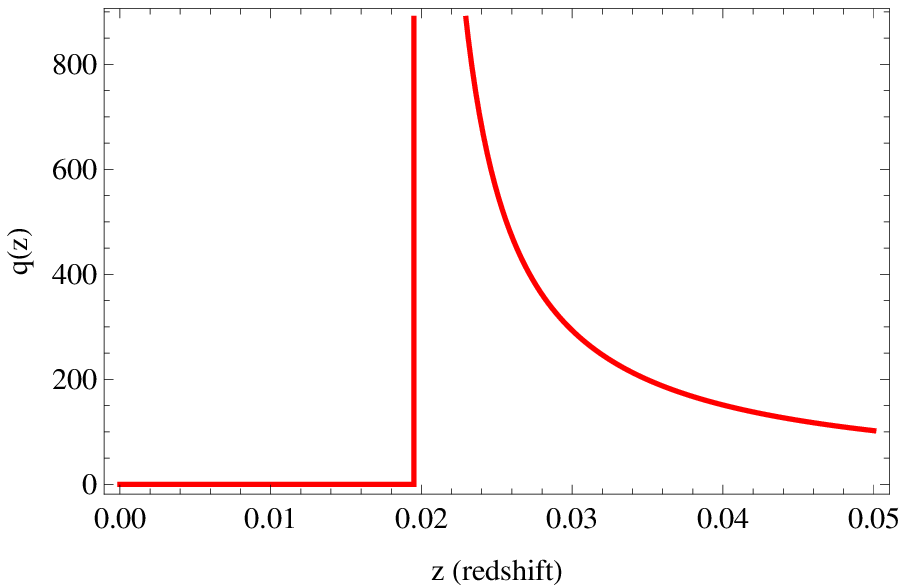}
\caption{The deceleration parameter $q(z)$ as a function of $z$
for the Einstein Gauss-Bonnet model with $\xi (\phi)=V_0 e^{\beta
\phi ^4}$ and $V (\phi)=V_0 e^{-\beta  \phi ^4}$. The left figure
corresponds to a close up of the $q(z)$ behavior near $z=0$ and
the right figure is the behavior for higher redshifts. It can be
seen that the deceleration to acceleration behavior occurs nearly
at $z_t=0.02$, and this crucially depends on the initial
conditions of $\phi'(0)$. For the two plots we used the initial
conditions $H(0)=1$, $\phi (0)=1100$ and $\phi'(0)=-10^{-6}$ and
also we assumed that $V_0=10$ and $\beta=0.01$, but the last two
variables do not affect crucially the late-time
evolution.}\label{plot3}
\end{figure}
In the previous section we demonstrated how the single canonical
scalar field quintessential inflation scenario is modified in the
context of a simple Einstein Gauss-Bonnet extension. What now
remains is to investigate the late-time properties of the Einstein
Gauss-Bonnet quintessential inflation scenario. To this end, we
shall make use of the redshift parameter, which is defined in
terms of the scale factor as $1+z=\frac{1}{a}$, where we have set
the value of the scale factor at present time equal to one, that
is $a(z=0)=1$. For the late-time evolution study we shall focus on
the behavior of the deceleration parameter $q(z)$ which in terms
of the Hubble rate is defined as follows,
\begin{equation}\label{decelerationparameter}
q(z)=\frac{1+z}{H(z)}\frac{d H(z)}{d z}-1\, ,
\end{equation}
and of the total effective equation of state parameter (EoS)
$w_{eff}$ which is defined below,
\begin{equation}\label{totaleffectiveeoes}
w_{eff}=-1+\frac{2(z+1)}{3H(z)}\frac{d H}{d z}\, .
\end{equation}
For the purposes of our study we shall investigate the behavior of
the above physical quantities as functions of the redshift, for
redshifts in the range $z=[0,10]$, which means that we look back
in our Universe's past for at least $12.6$ billion years. For the
$\Lambda$CDM model, the deceleration to acceleration transition
occurs for $z\sim 0.4$, given that $\Omega_{DM}=0.286$ and
$\Omega_{DE}=0.714$. The study we shall perform in this section is
purely numerical, so we shall express the gravitational equations
of motion as functions of the redshift $z$, so by using the
following transformation rules for the derivative,
\begin{equation}\label{derivativestrans}
\frac{d}{d t}=-H(z)(1+z)\frac{d}{d z}\, ,
\end{equation}
and also by assuming the presence of a dark matter fluid with
energy density $\rho_m=\rho_{m0}(1+z)^3$, the gravitational
equations of motion become,
\begin{align}\label{eqnmotion1}
& (z+1) H(z) \left((z+1) H'(z) \phi '(z)+(z+1) H(z) \phi
''(z)+H(z) \phi '(z)\right)-3 (z+1) H(z)^3 H'(z) \\ \notag & 24
\left(H(z)^4-(z+1) H(z)^4 H'(z)\right) \xi_{\phi}(\phi
(z))+V_{\phi}(\phi (z))=0\, ,
\end{align}
\begin{align}\label{eqnmotion2}
& -2 (z+1) H(z)^2 H'(z)-8 (z+1) H(z)^4 \xi_{\phi}(\phi (z)) \phi
'(z)+(z+1)^2 H(z)^2 \phi '(z)^2+\rho_{m0} (z+1)^3\\ \notag & -8
H(z)^2 \left( \xi_{\phi}(\phi (z)) \left((z+1) H(z) \left((z+1)
H'(z) \phi '(z)+(z+1) H(z) \phi ''(z)+H(z) \phi
'(z)\right)\right)+(z+1)^2 H(z)^2  \xi_{\phi \phi}(z) \phi
'(z)^2\right)\\ \notag & -16 (z+1) (z+1) H(z)^3 H(z)
\xi_{\phi}(\phi (z)) H'(z) \phi '(z)=0\, ,
\end{align}
where $\xi_{\phi}$, $\xi_{\phi \phi}$ and $V_{\phi}$ are defined
as follows,
\begin{equation}\label{defsinitiosn}
\xi_{\phi }=\frac{d \xi (\phi )}{d \phi },\,\,\,\xi_{\phi \phi
}=\frac{d^2 \xi (\phi )}{d \phi^2},\,\,\,V_{\phi }(\phi )=\frac{d
V (\phi) }{d \phi }\, ,
\end{equation}
and the primes in Eq. (\ref{eqnmotion1})-(\ref{eqnmotion2}) denote
differentiation with respect to the redshift $z$. The differential
equations  (\ref{eqnmotion1})-(\ref{eqnmotion2}) can be solved
numerically, so we shall perform a thorough analysis for both the
models  (\ref{quintpot1})-(\ref{xiphicoupling1}) and
(\ref{potentialxithrirdnew2})-(\ref{slowrollindicescase1new2}), by
using various initial conditions. Recall that the inflationary era
for both the aforementioned Einstein Gauss-Bonnet models occurs
for small values of the scalar field, so at late-times the scalar
field must take relatively large values. An examination of the
behavior of the solutions corresponding to the differential
equations (\ref{eqnmotion1})-(\ref{eqnmotion2}), reveals that a
deceleration to acceleration transition for small redshifts of the
order $z\sim 0.5$ can occur only if the ``velocity'' of the scalar
field $\phi' (0)$ is negative and small.

\begin{figure}[h]
\centering
\includegraphics[width=18pc]{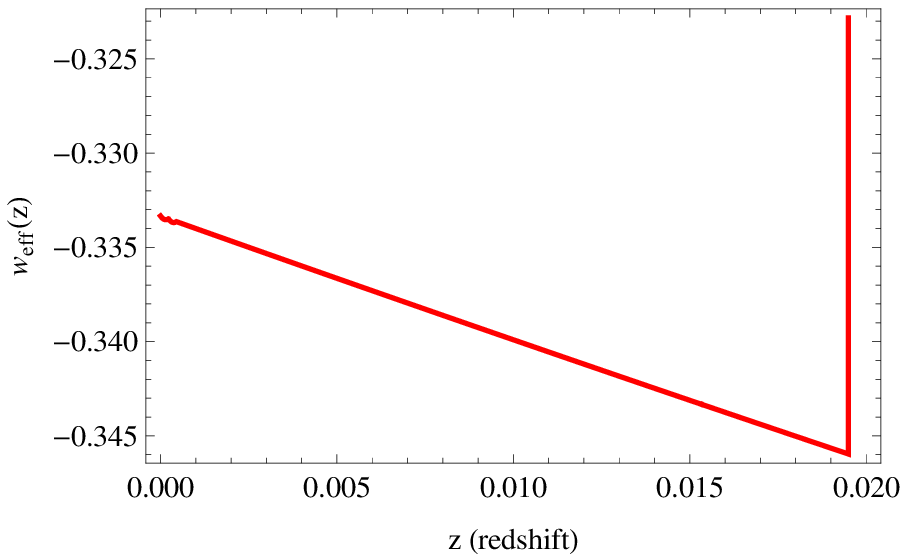}
\includegraphics[width=18pc]{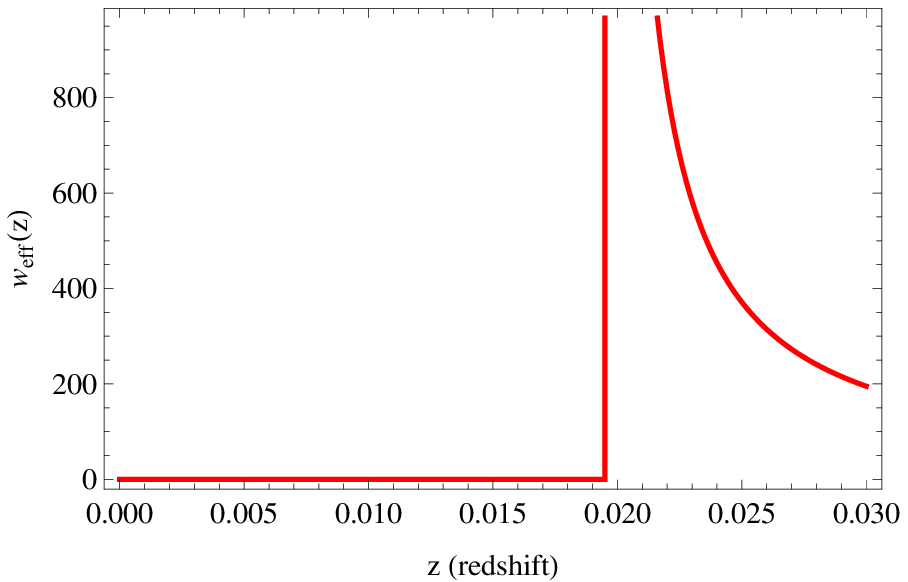}
\caption{The total equation of state parameter $w_{eff}(z)$ as a
function of $z$ for the Einstein Gauss-Bonnet model with $\xi
(\phi)=V_0 e^{\beta  \phi ^4}$ and $V (\phi)=V_0 e^{-\beta  \phi
^4}$. The left figure corresponds to a close up of the
$w_{eff}(z)$ behavior near $z=0$ and the right figure is the
behavior for higher redshifts. It can be seen that after $z=0.02$
the Universe is accelerating for some time.  For the two plots we
used the initial conditions $H(0)=1$, $\phi (0)=1100$ and
$\phi'(0)=-10^{-6}$ and also we assumed that $V_0=10$ and
$\beta=0.01$, but the last two variables do not affect crucially
the late-time evolution.}\label{plot4}
\end{figure}
\begin{figure}[h]
\centering
\includegraphics[width=18pc]{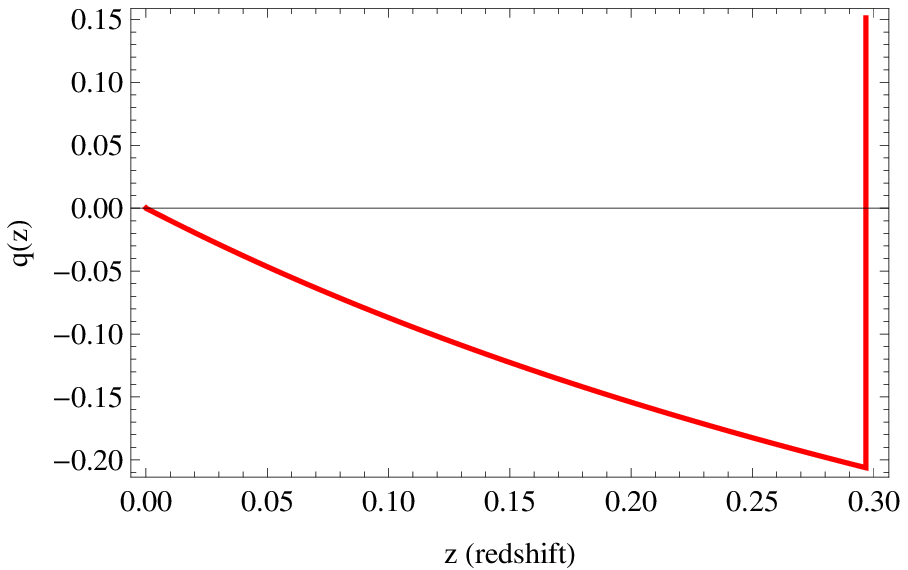}
\includegraphics[width=18pc]{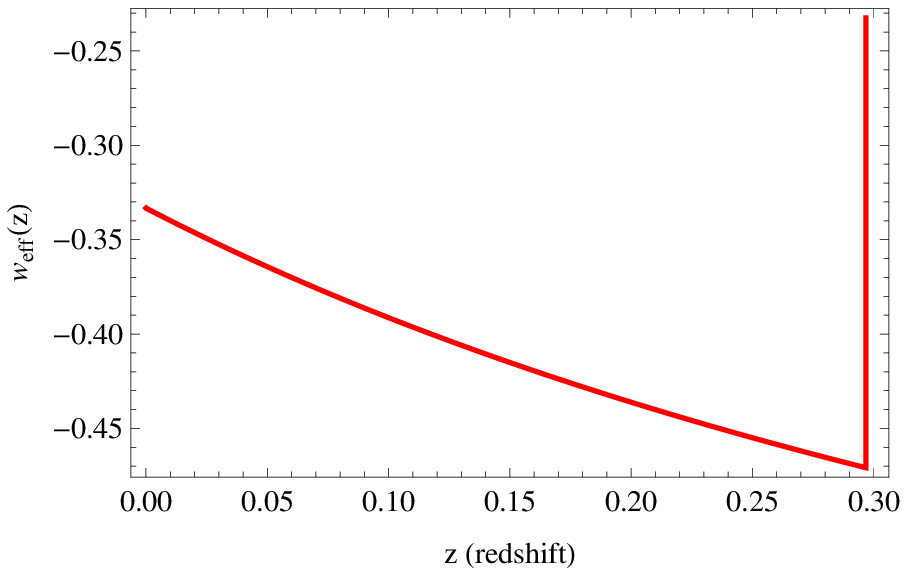}
\caption{The deceleration parameter $q(z)$ (left) and the
effective equation of state parameter $w_{eff}(z)$ (right), as
functions of $z$ for the Einstein Gauss-Bonnet model with $\xi
(\phi)=\frac{\beta}{\phi^4} $ and $V (\phi)=V_0 \phi ^4$. For the
two plots we used the initial conditions $H(0)=1$, $\phi
(0)=10^{-15}$ and $\phi'(0)=-10^{6}$ and also we assumed that
$V_0=1$ and $\beta=0.01$, but the last two variables do not affect
crucially the late-time evolution.}\label{plot5}
\end{figure}

If on the other hand $\phi' (0)>0$, the deceleration to
acceleration transition occurs for $z>20$ which is unacceptable
phenomenologically. For the model
(\ref{quintpot1})-(\ref{xiphicoupling1}), the behavior of the
deceleration parameter and of the total EoS parameter as functions
of the redshift is presented in Figs. \ref{plot3} and \ref{plot4}.
For all the plots we have used the values $\rho_{m0}=0.1$,
$V_0=10$, $\beta=0.01$ and the initial conditions $H(0)=1$, $\phi
(0)=1100$ and $\phi'(0)=-10^{-6}$. Also an investigation of the
behavior for various ``velocities'' of the scalar field at zero
redshift, indicates that as the absolute value of the velocity
drops, the redshift for which the deceleration to acceleration
occurs increases. In the plots, the deceleration to acceleration
transition occurs approximately at $z=0.02$, and this depends
strongly on the initial condition chosen for $\phi'(0)$. From a
phenomenological point of view, the behavior of the model at late
times indicates that it can describe a decelerating era until some
critical redshift is reached, at which point the Universe
super-decelerates. After that critical redshift, the Universe
accelerates in a decreasing rate until the present time era, in
which the deceleration parameter approaches slowly the value zero.
The same behavior occurs for the model
(\ref{potentialxithrirdnew2})-(\ref{slowrollindicescase1new2}), so
we omit it. There seems to be a pattern of same behaviors for the
Einstein Gauss-Bonnet models of the form $\xi (\phi )\sim
\frac{1}{V(\phi )}$, so we investigated another model of this
form, with $\xi (\phi)=\frac{\beta}{\phi^4} $ and $V (\phi)=V_0
\phi ^4$. In this case the slow-roll inflationary era occurs for
large values of the scalar field, so at late times the scalar
field should in principle take small values. In Fig. \ref{plot5}
we present the deceleration $q(z)$ (left) and the effective
equation of state parameter $w_{eff}(z)$ (right), as functions of
$z$ for the initial conditions $H(0)=1$, $\phi (0)=10^{-15}$ and
$\phi'(0)=-10^{6}$ and for $V_0=1$ and $\beta=0.01$. As it can be
seen in Fig. \ref{plot5}, the behavior of both $q(z)$ and
$w_{eff}(z)$  is quite similar to the previous quintessential
Einstein Gauss-Bonnet models we studied.
Thus we may conclude that the Einstein Gauss-Bonnet models with
$\xi (\phi )\sim \frac{1}{V(\phi )}$ seem to produce the same
phenomenology for late times, which indicates that the Universe
decelerates until some redshift, and then after a steep
deceleration point, an acceleration era occurs which has a
decreasing rate. This result is different though from the
$\Lambda$CDM model, which describes a nearly constant acceleration
rate until present time. Thus the resulting picture is not
compatible with the $\Lambda$CDM model, and although the early
phenomenology of the models we studied is quite compatible with
the observations, the late-time phenomenology is peculiar, however
an accelerating evolution is generated.

\section{Conclusions}

In this paper we studied the early and late-time evolution of the
Universe in the context of Einstein Gauss-Bonnet quintessential
models. With regard to the early-time behavior, we presented the
slow-roll formalism of the theory and we investigated if a viable
inflationary era can be achieved. As we demonstrated, the spectral
index and the tensor-to-scalar ratio corresponding to the models
we studied can be compatible with the observational data coming
from Planck and the BICEP2/Keck-Array data, for a wide range of
the parameter values. Actually we showed that the viability of the
quintessential models is enhanced in the context of the Einstein
Gauss-Bonnet theory, in comparison to the single canonical scalar
field case. With regard to the late-time era, all the models we
studied result to a decelerating era until some critical redshift,
at which point a super deceleration occurs, and eventually an
acceleration era follows. Notably, the rate of the acceleration
decreases until present time, and also the critical redshift at
which the deceleration to acceleration transition occurs,
crucially depends on the initial conditions chosen for the scalar
field. The behavior of the models seems to be the same for all the
models for which the scalar coupling to the Gauss-Bonnet scalar
$\xi (\phi)$ satisfies $\xi (\phi)\sim \frac{1}{V(\phi)}$. Also,
although a late-time acceleration era is produced for the
quintessential Einstein Gauss-Bonnet model we studied, the
evolution is different in comparison to the $\Lambda$CDM model. In
principle, the inclusion of higher order derivatives of the scalar
field can alter this behavior, so in a future work we aim to
examine this issue in more detail.

We need to note that what we tried to demonstrate here is the possibility to describe the inflationary and the dark energy eras, namely the two accelerating eras of our Universe, using the theoretical framework of Einstein-Gauss-Bonnet gravity. Such an idea is not new and it was firstly introduced in Ref. \cite{Nojiri:2003ft}, in the context of $f(R)$ gravity. However the difference is that in the Einstein-Gauss-Bonnet gravity, the behavior of the quintessential potential is peculiar and produces a super-decelerating era, absent in the context of $f(R)$ gravity. To our opinion the $f(R)$ gravity framework provides a much more solid phenomenological framework.

Finally, another question is whether the scalar field with such a quintessential potential used in this paper, can act as some dark matter component. This is a hard question to answer in brief, since up-to-date there is no evidence of particle dark matter, so perhaps in the context of some Chameleon theory of dark matter, this might be possible. Nevertheless the observational data seem to favor the particle nature for some or all of the components of dark matter, so we leave this question for a future work.

\section*{Acknowledgments}

Financial support by the Research Committee of the Technological
Education Institute of Central Macedonia, Serres, under grant
SAT/ME/130319-111/17, is gratefully acknowledged.

\end{document}